# Feasibility Study on Disaster Management with Hybrid Network of LTE and Satellite links.

Mboli Sechang Julius (MSc. PMSCOM, BEng. Elect/Elect)

*Abstract*- Our world today is highly vulnerable to either natural or artificial catastrophes and therefore, Public Protection and Disaster Relief (PPDR) operators need reliable wireless communications for successful operations especially in critical rescue missions. In most of such cases, PPDR dedicated or commercial terrestrial networks have always been used which at most times lead to unsuccessful operations. This is due to the fact this networks are all infrastructure-based which can be destroyed, fail to deliver the required service or the networks are not able to support and sustain the sudden traffic surge. Long-Term Evolution (LTE) is earmarked as the future candidate technology for PPDR purpose and so much effort have been put into it in terms of research to see how suitable architecture that will meet mission-critical requirements can be developed. This is based on the assumption that terrestrial networks will always be available. Unfortunately, in worst case scenarios, infrastructures might get damaged totally or might be destroyed by subsequent incidences. As a result, when critical situation as such occurs, adequate guarantees can only be possible in the hypothesis of very high financial involvement.
Fortunately, considering availability, coverage ubiquity and reliability, satellite technologies have lately proven to be good. So, to maximise the high channel performance of terrestrial networks and the availability and reliability of non-terrestrial networks, the solution lies in a hybrid system. It is on this ground that this work deals with the integration of LTE and satellite networks in both infrastructure-based and infrastructure-less topologies for PPDR purpose. It is aim at providing people trapped in disaster and field operators with a transparent accessibility, broadband performance and guaranteed coverage even when infrastructures are damaged. The requirements are defined and the model simulated. The network is able to provide network coverage, enhanced capacity and promised greater resilience.
*Keywords*; Emergency network, service requirements, user requirements, TETRA, TETRAPOL, PPDR dedicated networks, Satellites Communications.

## I. INTRODUCTION

The technologies that are currently used for public protection and disaster relief (PPDR) purposes especially around the European Countries for their operations on the field are the ones that still depend on the analogue Professional Mobile Radio (PMR) or Terrestrial Trunked RAdio (TETRA) and TETRAPOL technologies which are able to provide but heavy set of voice-driven services [1]. These voice services provided are very essential for PPDR operators especially when on initial stages of the rescue mission but regrettably they cannot support and sustain applications that demand high bandwidth data and it is these applications that are on the high and ever increasing demand from PPDR community. Moreover, the PPDR community do not have a common communication infrastructure for the various entities like the ambulance services, fire departments and police corps. Therefore, for coordination purpose and high data oriented applications, the operators have to depend on commercial terrestrial networks [2].

In need for reliable and high communication systems for PPDR purpose, the next generation public safety network will still be LTE. These proposals which are either dedicated, hybrid or dedicated commercial solutions are not very sufficient and reliable in worst case scenarios where infrastructure-based terrestrial system are partially or totally destroyed by the disaster. This is true even when the system is supported by mobile ad hoc network that still rely on the terrestrial infrastructure-based system because when natural or artificial disaster occurs, network infrastructures may be totally destroyed which will cut off the entire area of the incident from network access. If probably, the incident left the area with terrestrial networks with partially damaged infrastructures, the increased network traffic in the site due to the incident will greatly affect communication for both rescue workers and the civilians or subsequent disaster will destroy the remaining infrastructures when it occurs in series thereby causing total absence of the network in the incident area. Though transmission quality and coverage can also be increased by LTE using spectrum holes through the concepts of cognitive radio access [3] but these does not still guarantee availability and resilience under catastrophes scenarios.

This work considers the employment of a new proposed hybrid network system that is based on both infrastructure-less and infrastructure-based approach to provide high bandwidth data-oriented applications through LTE base stations deployable during rescue mission backhauled on satellites system [2]. This system can provide coverage in situations where infrastructure-based systems are totally destroyed by the disaster without requiring civilians and even PPDR operators to use any special user equipment (UE) before accessing connectivity to the network. However, if the disaster left infrastructures partially functional like the case of Nepal earthquake in April 2015 [4], then, the deployed infrastructure-less service should provide

PPDR operators with broadband connection, high resilience and extended LTE coverage [2].

## II. DESIRED APPLICATIONS FOR DISASTER MANAGEMENT

The analysis as geared toward the identification and classification of applications and services needful for disaster relief as discussed in public Protection and Disaster Relief Transformation Centre (PPDR-TC) [5] produces five unique categories of communications. These services and applications are further categorized into common requirements and specific requirements. The two specific requirements are voice and data communications while the rests are classed as common requirements as outline here [1];

(a) Voice communication; This include all extended features such as push to-talk (PTT), direct mode operation (DMO) and group calls are of utmost important to emergency scenarios.
(b) Narrowband data communication which can be used for instance in messaging services.
(c) Broadband data communications which is specifically needed for transmission and reception of large data such as images and video files.
(d) Video streaming which is more critical in terms of latency though it is still similar to broadband data.
(e) Non-standard communication which may include but not limited to air-to-ground communications from helicopters and communications in enclosed and remote areas where extension of network is needed with the aid of repeater stations,

Video transmission have always been performed through commercial terrestrial networks where the infrastructures still exist and if destroyed or not functional, then mobile repeater station could be used which have also proven useful where applied.

## III. LOOPHOLES IN TECHNOLOGIES FOR EMERGENCIES

There are different network systems used for PPDR purposes and this includes PPDR-dedicated networks, commercial terrestrial networks and LTE which is earmarked as the future candidate for PPDR operations. This section presents the various gaps and challenges exhibited by these technologies at one time or another as were/are used by PPDR agencies for emergency operations.

*1. Emergency Dedicated Networks*

TETRA and analogue PMR are the networks used for PPDR communications in Europe currently. A data rate of about 28.8 Kbit/s is possible with TETRA though is not even deployed in all European Countries and TETRA2 is able to offer up to 400 Kbit/s which is only deployed in some local regions while some other Countries are still on the waiting list [1 6]. It is also a fundamental issue that current PPDR dedicated networks suffer deficient capabilities and limited spectrum harmonization. Data and video transmission and the effective QoS as needed by PPDR professionals still remains a major gap as these technologies are unable to offer high-speed data transmission. Analogue PMR systems are almost going out of their life cycle due to the fact that they are limited in terms of security and specific voice services though in some conditions, where signal properties degrade significantly, analogue PMR systems may be preferred over digital systems, because of their gradual degradations behaviour and therefore "cliff edge" effect (sudden loss of signal reception) in digital systems are avoided.

Remote areas not covered by terrestrial networks or when a backup transmission link is necessary in some situations, have been covered by Geosynchronous Orbit (GEO) satellites but the major problem of this system is their high altitude from the earth surface resulting in high propagation delay making call setup times longer and delayed voice acknowledgement. This kind of network is good but cannot still be sufficient by itself for critical scenarios.

*2. Commercial Terrestrial Networks*

Commercial terrestrial networks used for PPDR data-oriented communications in the European region are: GSM/GPRS/EDGE, CDMA2000, UMTS, HSPA and HSPA+ which has exhibited different limitations in different levels and at different events [1, 7, 8]. Though these technologies are able to provide higher bandwidth in standard conditions compared to the current dedicated PPDR networks, they may be congested when under emergency rescue situations and they have limited interoperability with other networks, moreover, they do not have DMO capabilities and are also incapable of conveying data intensive traffic. Because of probable end-to-end encryption by some users, there is no group calls or sufficient push-to-talk (PPT) implementation. A delay of about 5 seconds call setup time is also possible with some commercial terrestrial networks [9].

*3. Future Candidate Technologies*

The Consortium have listed LTE, IEEE 802.11, Satellite-over-IP, Worldwide interoperability for microwave access (WiMAX) and Mobile ad hoc Networks (MANETs) as future technologies for PPDR systems [1, 2, 10]. WiMAX, MANETS and IEEE 802.11 networks employ high frequency bands which limit their coverage range greatly though MANETs can avoid this limit by employing multi-hop transmission. Mission critical voice services such as PTT, DMO and group calls are natively not available with these networks since their IP-based nature only offers voice over IP (VOIP) calls [11].

In the case of LTE, these same issues exist, except for the latest releases [1, 12] which have standard that introduces some features that are specifically reserved for emergency services. These specific features are meant to enable proximity services and group calls.

WiMAX and IEEE 802.11 networks used shared licensed-exempted spectrum (common issue at high frequency bands) which may likely affect their connections availability. IEEE 802.11 networks are more affected in urban areas because many systems are already deployed for personal and business usage than they are in suburban or rural areas. There is also great dependency on mobility scheme whether high or low rate, number of users competing for access and distance from access points by IEEE 802.11 networks. It is possible to increase the overall coverage or bandwidth and reliability in MANETs by adding more devices to the networks but this result in high MANETs routing protocol overhead which consequentially lead to low performance again. Some limitations like connectivity models and network topologies are still found in LTE and it has even been proven possible that LTE base stations can be jammed with limited expenses and may not require any particular knowledge. The issue of long propagation delay still affect GEO Satellite-Over-IP systems due to its distance from the Earth. These propagation delays are actually very high especially when standard communication protocols are employed for terrestrial networks.

## IV. INTEGRATION OF SATELLITES TECHNOLOGIES

Satellite communications have recoded great successes in the past and still waxing strong and stronger due to its wide area coverage, rapid development, ubiquitous availability, reliability, security and privacy, flexibility and expandability, multi-cast content distribution and their speed to market for new services [13].The trend in fixed-satellite service (FSS) is the ability to support high frequency, capacity, broadband and IP oriented development, while High Definitions Television (HDTV) and multimedia services are possible with broadcasting satellite service (BSS). There are also new bands with improved spectral efficiency in mobile satellite services (MSS) and remote sensing is achievable with Earth observation and image processing technologies [14, 15].

However, with all the outlined advantages and latest trends in satellite communications in the preceding paragraph, satellite communications still suffers major disadvantages such as latency (RTT can be very high depending on the satellite employed), cost, scarcity of resources and high attenuation in some cases and these challenges will largely affect the operation of PPDR professionals especially in disaster relief aspect. Therefore, to exploit the advantages and current trend, the solution lies in integration with networks that will make up for the gaps.

*1. Integration of Satellite and Terrestrial Networks*

Terrestrial networks provide very good coverage in urban areas and in fact, the mature technology that is widely used, meeting most of the communications needs for the present and future generation of technology [15, 16]. Almost all terrestrial networks today are IP-based capable of delivering broadband communications. The trend in terrestrial networks is the target of machine to machine (M2M) communications, cloud computing and good network quality and security which is very important to PPDR communications.

However, using terrestrial networks only will not meet all the user and service requirements for PPDR applications as outlined in the preceding section because of the gaps in some aspects. Coverage is limited as remote areas are not covered or costly to cover and not economical to even do such projects. Resilience is not also guaranteed in terrestrial networks as it difficult to recover from damages due to major natural or man-made disasters.

Therefore, the idea of integration of satellite and terrestrial networks to exploit the advantages of both and as well combat the disadvantages of each together is the way forward for PPDR applications. It can be seen that the requirements of Terrestrial networks is to be ubiquitous, economical, high performance, high capacity and full coverage and the solution lies in satellite integration [17].

*2. Integration of Satellite and LTE*

LTE and LTE-A are chosen as the terrestrial networks to integrate with satellite and be able to provide applications because these are the current trending technologies capable of meeting PPDR applications requirements. If it is assumed that terrestrial networks are all destroyed in a disaster scenario, PPDR operators should still have their LTE coverage with aid of deployable mobile LTE repeater station backhauled on the satellite as seen in the system architecture of Figure 1. In most disaster situations, PPDR operators always choose either satellite or terrestrial networks but in this case, a hybrid network is proposed so that at all points, field operators have connectivity either by the usual

LTE network or by the ad hoc mobile LTE repeater stations (MEOC).

## V. PROPOSED SYSTEM ARCHITECTURE

The experiences from several catastrophes globally have shown that network infrastructures can be affected badly in worst-case scenarios where not only base stations are destroyed but other network components like link aggregation and even main infrastructures might be damaged alongside [2, 4]. Therefore, the proposed system here is to be a solution to such a terrible disaster scenario where deployment of infrastructure-less network is inevitable. If the terrestrial infrastructure-based networks are damaged totally, the deployed mobile ad hoc network will be the only source to provide coverage in the incident area. On the other hand, if only part of the terrestrial networks is affected so that capacity is reduced and traffic increased, the deployed infrastructure-less network will serve the purpose of increasing the network capacity and still make communications effective for both first responders and civilians in scene area.

The LTE being the future candidates' technology for PPDR and its ability to provide both narrowband and broadband communications is considered to be the main access technology while the satellites are to serve the purpose of backhaul as shown in Figure 1.The system indicates that LTE base stations are covered by the satellites backhaul to the mobile ad hoc system. In a disaster scenario, the priority should be first given to FRs who are connected to the deployed mobile network that serve as Mobile Emergency-controlled Centre (MEOC).

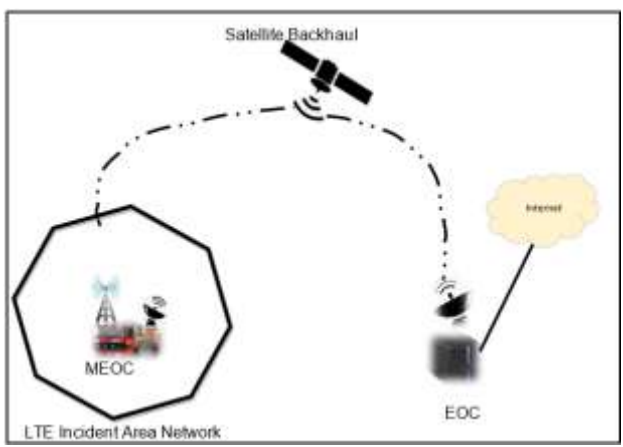

Figure 1: Proposed system Architecture

So the MEOC is able to effectively provide Incident Area Network (IAN) for FRs at mission field and thus serve as LTE mobile repeater station. The MEOC will need to be connected to Emergency Operation-control Centre (EOC) and achievable by using the reliable MEO satellite link as depicted in the system architecture **Figure 1** that is the only means the MEOC and EOC can establish communication link between them.

It is obvious that in the proposed system architecture, the Incident Area Network (IAN) is the infrastructure-less and mobile network that can only be deployed when catastrophe strike an area or a region. Accessibility to network on the incident location can be of three major cases as follows;

> ➢ No communications at all due to inability of persons to connect to either MEOC or terrestrial station.
> ➢ Possible communication if persons can connect to either MEOC or terrestrial station.
> ➢ Reliable communications as persons can connect to both MEOC and terrestrial bas station.

The three cases outlined above is not likely to be fixed or remain the same throughout the whole period of the rescue operations since a user can shift between them at any time especially when there is increase or decrease in the traffic of the terrestrial network, or a series of disasters where subsequent disaster destroys the remaining infrastructures or when the MEOC moves from its initial location and user is left out of MEOC coverage. The main aim of the scheme is that the first case is always avoided throughout the entire duration of the field operations.

The network architecture on a MEOC is illustrated [2]. The main subsystems of LTE; evolved packet core (EPC) and E-UTRAN are effectively integrated so the MEOC is able to supply IP connectivity to UEs since all the network components are IP-based. When the need for data arises, it can be routed through the integrated satellite link. From its architecture, it can be seen that the MEOC can function independently in the absence of any terrestrial core networks and it is therefore capable of avoiding any point of failure especially when multiple MEOCs are deployed.

*1. Provision of Service*

For the service provided of in a disaster area, the affected people, that is the people in distress should be able to connect to the commercial terrestrial networks and communicate their situations and location to their friends and loved ones and to contribute their part in rescue operations to rescue workers. As discussed in previous section, if these terrestrial networks are unable to provide connectivity to them due to total damage or destructions or increased traffic, they must connect to MEOC which will equally provide LTE service. Usually, in disaster scenarios, FRs have often relied on these commercial terrestrial networks [18] and this will even congest the network the more except when prioritization mechanisms are employed so it is best to keep this method as the last option. The affected people may be trapped, injured or even facing death and so at least voice and text messaging connectivity should be provided for them while FRs should have additional services, particularly broadband data support, specific voice capabilities and text messaging.

The recent LTE standard releases do not really guarantee provision of special services such as DMO and Group calls [2, 19], there may be need for additional services and external platforms like the IP Multimedia subsystems as defined by Open Mobile Alliance (OMA) to offer advance voice services through LTE. For broadband requirement, it is necessary to dedicate a channel via the MEOC with high LTE priority that will provide broadband data capabilities for FRs and excess MEOCS network capacity with low priority can be open for people in distress. So, it is very clear that MEOC network serve two purposes: apart from providing PPDR dedicated channel to FRs, it also provide people in distress with additional network capacity. LTE bearer services can be utilized extensively here to achieve this prioritization. A bearer can be defined as an IP packet flow that specifies particular QoS between a UE and a gateway and with this concept, a user can work with multiple bearers at the same time; For instance, it is possible that a FR while doing a file upload, can still be making emergency voice call (VOIP) concurrently [20]. A best effort bearer (UDP) is required for the file upload while a VOIP bear with dedicated resources is crucial QoS for voice call. With this concept, since once a UE connects to LTE network, is assigned an IP address via the P-GW interface, it is possible to assign best effort bearer to civilians by default and while dedicated bearers goes to FRs. The IP address allocated by P-GW to UE is based on the subscription data obtained from HSS and this ensures that one default bearer is always established at a time.

2. Discussion On Comparing Amplify and Forward (AF) And Decode and Forward (DF) For PPDR

In this section, a comparison of Amplify and Forward (AF) and Decode and Forward (DF) protocol with direct path will be demonstrated to show performance of the three different protocols in a Two-Way Radio Network (TWRN) of cooperative communications. Most TWRN communications happens in three time slots, so if node a (assumed to be UEs in this case) is able to send packet to both node b (can be terrestrial LTE eNodeB or MEOC eNodeB) and node c (Satellite) as illustrated in Figure 2. In the first time slot, $s_a[k]$ for $k = 0,1,…,K_a - 1$ can be used to represent the information bits in a block from node a. By performing channel coding with interleaving and binary shift keying (BPSK) with (M = 2 as the modulation order), the output BPSK symbols is now denoted as $x_a[n]$ for $n = 0,…,N - 1$ where N is the length of the transmission block.

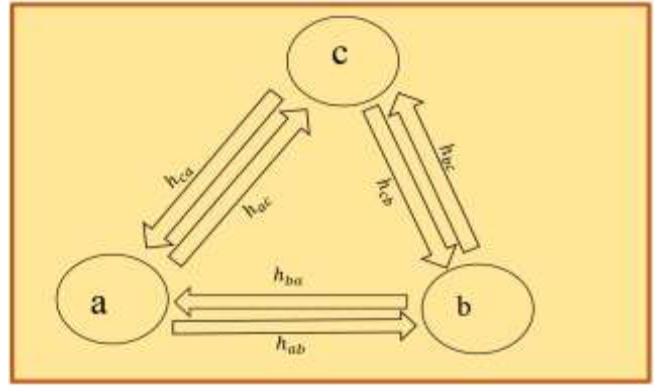

Figure 2: Comparison of AF and DF in TWRN.

Therefore, node b and node c will have the following received baseband signal respectively;

$$r_{ab}[n] = h_{ab}[n]\sqrt{P_a}x_a[n] + v_{ab}[n] \quad (1)$$

$$r_{ac}[n] = h_{ac}[n]\sqrt{P_a}x_a[n] + v_{ac}[n] \quad (2)$$

Where $P_a$ represents the power which node a transmits with, $h_{ab}[n]$ is the channel response while the AWGN noise at node b is denoted by $v_{ab}[n]$ with $CN(0,\sigma^2)$ as its distribution. It is a general principle in block fading channel model that length of the codeword spans various coherence periods, the entire codeword can be considered to be broken into several L blocks with Q as the approximate block length which is also equal to the channel coherence time. Therefore, about Q symbols are covered within the channel coherence time, so that N = QL. For the codeword to undergo slow fading channel, $h_{ab}[n]$ need to be uniform for all values of $n$. Similarly, $h_{ac}[n]$ and $v_{ac}[n]$ are the corresponding channel response and noise components from node a to node c with the assumption that all the nodes have same power even though this is not practically possible.

For when node b sends its packets to node a and node c in the second time slot, Let $s_b[k]$ for $k = 0,1,…,K_b - 1$ be the information bits block from node b. The result of channel coding with interleaving and BPSK modulation of $s_b[k]$ will produce $x_b[n]$ for $n = 0,…,N - 1$ as the output BPSK symbol with length of transmission block equals N. It is essential to note here that $K_a$ and $K_b$ may not be the same and node a and node b may not necessarily utilize the same coding scheme. For the purpose of this work and for convenience, N as the length of the packet is assumed to be same for both node a and node b and even node c. In real life practice, N can be made equal at various nodes by applying puncturing techniques which is simply puncturing some bits that are coded if the length of the codeword at one node is larger than N [21]. The received signals at node a and node c as node b sends in this second time slot are respectively;

$$r_{ba}[n] = h_{ba}[n]\sqrt{P_b}x_b[n] + v_{ba}[n] \quad (3)$$

$$r_{bc}[n] = h_{bc}[n]\sqrt{P_b}x_b[n] + v_{bc}[n] \quad (4)$$

The definition of channel responses and noises are similar to those of equations (1) and (2) while $P_b$ is the transmit power of node b.

With two-way traditional network coding, the relay node which can be node c in this case decodes the packets which it received in the first and second time slots and verify that these packets are decoded correctly, it then perfectly regenerates BPSK symbols from node a and node b, that is $x_a[n]$ and $x_b[n]$ respectively and Dot operation of the BPSK symbols which is the simplest network coding [22] is now carried out as follows;

$$x_c[n] = (x_a[n].x_b[n]) \quad (5)$$

The BPSK symbol $x_c[n]$ is forwarded from node c to node a and to node b after performing network coding on it in the third time slot. Node a and node b will respectively receive the following signals from node c.

$$r_{ca}[n] = h_{ca}[n]\sqrt{P_c}x_c[n] + v_{ca}[n] \quad (6)$$

$$r_{cb}[n] = h_{cb}[n]\sqrt{P_c}x_c[n] + v_{cb}[n] \quad (7)$$

Where $P_c$ denote transmit power of node c while channel response and noises are defined in similar manner to (1) and (2).

This network coding scheme is based on the assumption that a special UE can also be employed to communicate with the satellite directly and may also communicate with satellite through the LTE network provided by MEOC or terrestrial LTE. It should be noted that this scheme is still limited in terms of diversity and path gains in that if the both packets cannot be simultaneously and correctly decoded, node c is just going to broadcast the packet decoded correctly without performing network coding. If on the other hand, both packets fail to be decoded correctly, then nothing will happen in the third time slot. A more efficient scheme could be used which is a hybrid AF and DF scheme [23] but it is beyond the scope of this work for now. The result of this graphical comparison between AF and DF is given in Figure 3

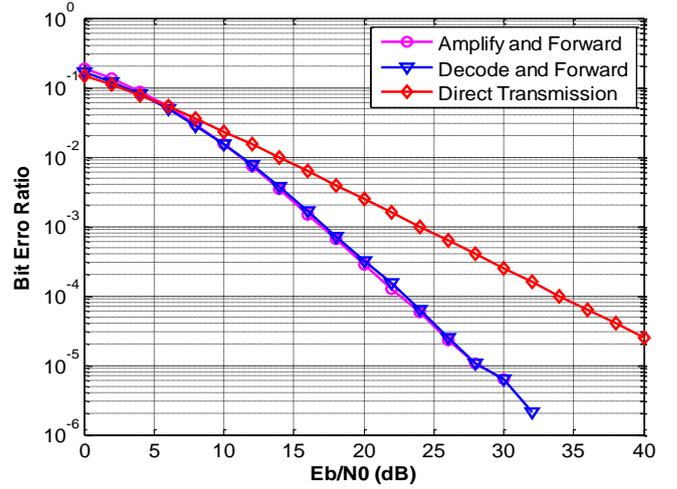

Figure 3: Comparison of AF, DF and Direct path

## VI. CONCLUSION

The usage of infrastructure-based terrestrial networks by PPDR operators in disaster scenarios and major events is vulnerable to failure due to several reasons including destruction of infrastructures by the catastrophe and probable traffic surge. To address this situation, the consortium after critical analysis, came up with the most hopeful network approach which is a hybrid network of satellites, PPDR dedicated networks and terrestrial networks especially LTE. To that effect, the proposed network architecture presented in this work integrates satellites and LTE for PPDR purpose. The proposal is based on a deployable mobile technology that is capable of extending LTE coverage to disaster area employing the satellite as backhaul medium with LTE as the access network. This is achievable by deploying multiple mobile units that ensures easy connectivity and guarantees high performance and resiliency. In addition, with just simple configuration, the architecture is capable of providing service with both infrastructure-less and infrastructure-based topologies. This network scheme increases reliability of the existing commercial and PPDR dedicated network during impromptu incidents and at the same time leverages terrestrial infrastructures. The scheme also proposed the possibility of solving interoperability issue with legacy cellular technologies to ease operations at international borders with the potential of providing communications in tunnel and indoors when under disasters scenario.

Simulations were carried out using wireless cooperative network with amplify and forward protocol to model the different channels of the proposed network architecture in MATLAB, using Rayleigh fading channel for UE and the mobile unit model (MEOC) while Rician channel was used for Earth station (EOC) and satellite link channel. The channel performance of the networks was evaluated with graph of bit error rate and energy per bit per spectral noise density.

Conclusively, the proposed network is seen to offer field operators with broadband capabilities, high performance capacity, and resilient network with good reliability and ability to extend coverage with easy connectivity and transparent handovers even when terrestrial infrastructure-based network are absent.

## VII. RECOMMENDATIONS AND FUTURE WORKS

The proposed network is designed to be of great reliability, high resiliency, extended coverage guaranteed performance with regard to QoS and broadband capability, however, with the limitations in the previous section, there is still need for more works and recommendations as follows;

- The use of various channel models to verify the efficiency of the network.
- The possible modification of the scheme to include missing critical voice services such as DMO, PTT and group calls by possibly integrating PPDR systems like TETRA and TETRAPOL into the scheme.
- The use of other network design and simulation tools that will give possibility of designing the complete hybrid network to evaluate channel performance with respect to change in traffic with both X2 and S1 interface handovers.
- To design a model that can be able to evaluate mobility in detail as it is a major requirement of PPDR operators with their nature unpredictable movement.
- To use diverse channel models to mimic the effect of very hot environments under fire, collapsed structures and buildings and other critical emergency scenarios to be able to evaluate and investigate their effects on communication systems.
- To use other channel model to simulate the principle of adaptive bit modulation as part of measure for assigning QoS since that is what is done in LTE today.
- The use of Low Altitude Platforms (LAPs) and Remote Radio Heads (RRHs) might also be necessary where incidents locations become impracticable for MEOC deployment due to inaccessible remote terrain.

## VIII. REFERECES

[1] C. A. Grazia, M. Klapez, N. Patriciello, M. Casoni, A. Amditis, E. Sdongos, H. Gierszal, D. Kanakidis, C. Katsigiannis, K. Romanowski, P. Simplicio, A. Oliveira, S. Sonander, and J. Jackson, "Integration between terrestrial and satellite networks: the PPDR-TC vision," in *Wireless and Mobile Computing, Networking and Communications (WiMob), 2014 IEEE 10th International Conference on*, 2014, pp. 77-84.

[2] M. Casoni, C. A. Grazia, M. Klapez, N. Patriciello, A. Amditis, and E. Sdongos, "Integration of satellite and LTE for disaster recovery," *Communications Magazine, IEEE*, vol. 53, no. 3, pp. 47-53, 2015.

[3] T. F. Rahman and C. Sacchi, "Opportunistic radio access techniques for emergency communications: Preliminary analysis and results," in *Satellite Telecommunications (ESTEL), 2012 IEEE First AESS European Conference on*, 2012, pp. 1-7.

[4] D. Madory. Earthquake rocks Internet in Nepal [Online]. Available: http://research.dyn.com/2015/04/earthquake-rocks-internet-in-nepal/ [Accessed: 09/06/2015].

[5] P. Simplício, J. Belfo, and A. Oliveira, "PPDR'S TECHNOLOGICAL GAPS," 2013.

[6] M. Redman, "PMR Technology Appendix," in *Digital Project Report*, Version 1 ed, 2002.

[7] "User requirements and spectrum needs for future European broadband PPDR systems (Wide Area Networks),," 20-24th May, 2013.

[8] L. Goratti, G. Steri, K. M. Gomez, and G. Baldini, "Connectivity and security in a D2D communication protocol for public safety applications," in *Wireless Communications Systems (ISWCS), 2014 11th International Symposium on*, 2014, pp. 548-552.

[9] S. Borkar, D. Roberson, and K. Zdunek, "Priority Access for public safety on shared commercial LTE networks," in *Telecom World (ITU WT), 2011 Technical Symposium at ITU*, 2011, pp. 105-110.

[10] D. Munir, G. Jaheon, and C. Min Young, "Selection of UE relay considering QoS class for public safety services in LTE-A network," in *Communications (APCC), 2014 Asia-Pacific Conference on*, 2014, pp. 401-405.

[11] K. C. Budka, T. Chu, T. L. Doumi, W. Brouwer, P. Lamoureux, and M. E. Palamara, "Public safety mission critical voice services over LTE," *Bell Labs Technical Journal*, vol. 16, no. 3, pp. 133-149, 2011.

[12] T. Doumi, M. F. Dolan, S. Tatesh, A. Casati, G. Tsirtsis, K. Anchan, and D. Flore, "LTE for public safety networks," *Communications Magazine, IEEE*, vol. 51, no. 2, pp. 106-112, 2013.

[13] B. Evans, M. Werner, E. Lutz, M. Bousquet, G. E. Corazza, and G. Maral, "Integration of satellite and terrestrial systems in future multimedia communications," *Wireless Communications, IEEE*, vol. 12, no. 5, pp. 72-80, 2005.

[14] P. J. L. F, "Convergence of Terrestrial and Satellite Communication networks " 2014.

[15] R. E. Sheriff and Y. F. Hu, *Mobile satellite communication networks*. New York; Chichester: Wiley, 2001.

[16] S. Yadav. Convergence of Terrestrial and Satellite mobile communication systems [Online]. Available: https://sachendra.wordpress.com/2009/04/23/convergence-of-terrestrial-and-satellite-mobile-communication-systems/ [Accessed: 22/07/2015].

[17] G. Iapichino, C. Bonnet, O. del Rio Herrero, C. Baudoin, and I. Buret, "A mobile ad-hoc satellite and wireless Mesh networking approach for Public Safety communications," in *Signal Processing for Space Communications, 2008. SPSC 2008. 10th International Workshop on*, 2008, pp. 1-6.

[18] Andre Oliveira and P. Simplício, "PPDR's Needs and Requirements D2.2," 01/07/2013 2013.

[19] D. Astely, E. Dahlman, G. Fodor, S. Parkvall, and J. Sachs, "LTE release 12 and beyond [Accepted From Open Call]," *Communications Magazine, IEEE*, vol. 51, no. 7, pp. 154-160, 2013.

[20] J. Korhonen, *Introduction to 4G mobile communications*. Boston: Artech House, 2014.

[21] A. Burr, "Turbo-codes: the ultimate error control codes?," *Electronics & Communication Engineering Journal*, vol. 13, no. 4, pp. 155-165, 2001.
[22] S. Riedel, "Symbol-by-symbol MAP decoding algorithm for high-rate convolutional codes that use reciprocal dual codes," *Selected Areas in Communications, IEEE Journal on*, vol. 16, no. 2, pp. 175-185, 1998.
[23] Z. Yu, W. Xinhua, and Z. Tianshi, "Hybrid AF and DF with network coding for wireless Two Way Relay Networks," in *Wireless Communications and Networking Conference (WCNC), 2013 IEEE*, 2013, pp. 2428-2433.

**Biography**

Mboli Sechang Julius graduated with Distinction in MSc Personal, Mobile and Sattlite Communication from the Faculty of Engineering and Informatics, University of Bradford, United Kingdom December 2015. He had BEng Electrical and Electronics Engineering from Modibbo Adama university of Technology, Yola, Adamawa State, Nigeria.